# A Configurable Memristor-based Finite Impulse Response Filter

*Mohammad Hemmati, Vahid Rashtchi* [*]*, Ahmad Maleki, Siroos Toofan*
*Department of Electrical Engineering, University of Zanjan, Zanjan, Iran*
[*]Corresponding Author: rashtchi@znu.ac.ir

*Abstract*—**There are two main methods to implement FIR filters: software and hardware. In the software method, an FIR filter can be implemented within the processor by programming; it uses too much memory and it is extremely time-consuming while it gives the design more configurability. In most hardware-based implementations of FIR filters, Analog-to-Digital (A/D) and Digital-to-Analog (D/A) converters are mandatory and increase the cost. The most important advantage of hardware implementation of a FIR filter is its higher speed compared to its software counterpart.**
**In this work, considering the advantages of software and hardware approaches, a method to implement direct form FIR filters using analog components and memristors is proposed. Not only the A/D and D/A converters are omitted, but also using memristors avails configurability. A new circuit is presented to handle negative coefficients of the filter and memristance values are calculated using a heuristic method in order to achieve a better accuracy in setting coefficients. Moreover, an appropriate sample and delay topology is employed which overcomes the limitations of the previous research in implementation of high-order filters. Proper operation and usefulness of the proposed structures are all validated via simulation in Cadence.**

## 1. Introduction

After digital signal processors were introduced to the market, their unique features excelled analog signal processing methods; according to the Moore's law, processors become more complex and faster every year. In addition, their programmability gives the designer more flexibility. This trend has made digital signal processors more popular and it has led the analog signal processing units become rare[1, 2]. In recent years, increasing speed and reducing size of transistors have been a challenge. Thus, scientists have been looking for any new methods to enhance the performance of processors independent of the so-called trend. One of the approaches is using analog circuits besides digital signal processing units to overcome limitations of digital circuits [3]. This scheme attempts to build or enhance digital structures which not only their execution within digital processors consumes too much energy, but also they can be implemented using pure analog circuits. Filtering is an example of such operations which it is ubiquitous in signal processing applications and it could be a burden on processors. There are various methods to implement filters; practical and well-known topologies are

Infinite Impulse Response (IIR) and Finite Impulse Response (FIR) [4] while this work focuses on FIR filters, for they are widely used in signal processing applications. Structure of a direct form FIR filter for $m$ samples is shown in Fig. 1. $b_i(i = 0, 1, 2, \dots, m)$ are filter coefficients; $x[n]$, as the $n$-th sample, passing through each $z^{-1}$, is delayed for one sampling period. $y[n]$, is the output of filter which is determined by (1).

$$y[n] = b_0 x[n] + b_1 x[n-1] + \cdots + b_m x[n-m] \qquad (1)$$

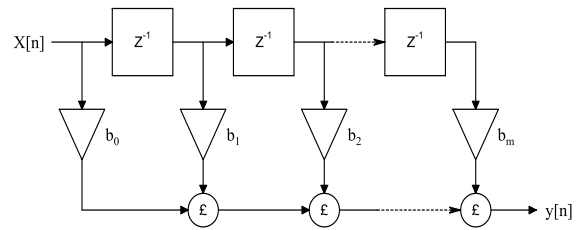

**Fig. 1** *Direct form FIR filter*

### 1.1. Evolution of Memristors

Memistors were devised in 1960 by Widrow as a 3-terminal element to be used in adaptive circuits; its resistance was adjustable by one of its terminals which receives DC current [5] . However, the introduced memistor could not act as the fourth circuit theory element as its I-V characteristic was not fixed and it could only be determined within the circuit which it is used in [6]. Accordingly, in 1968, Fano and colleagues claimed that except resistor, inductor, and capacitor, there is another missing circuit element [7]. In 1971, an article titled "Memristor- The Missing Circuit Element" proposing the fourth 2-terminal circuit element was presented by Chua. He believed that except three main circuit elements (resistor, capacitor, and inductor), another element called "memristor" exists which there is a function between its current flow and its conductance [8]. In 2008, researchers in HP laboratories implemented memristors physically [9] which their circuit model for a memristor is shown in Fig. 2. The black colored end specifies the polarity of the memristor. If the current flowing through the memristor is injected into it via this end, the resistance of the device is decreased and if the polarity of the current is reversed, the resistance would be increased [10]. Many different models have been presented for memristors [11–18]. In this work, the HP model is utilized. Today, memristors are used in several applications such as programmable analog circuits [19, 20], digital circuits [21], accelerators [22, 23], signal processing [24, 25], image processing [26] , neural networks[22, 27, 28], and etc.

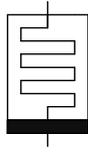

**Fig. 2** *Circuit schematic of a memristor*

### 1.2. *Background of Memristor-based FIR Filters*

There are many different methods to implement FIR filters [29–32], but a new approach has been proposed by [33] , shown in Fig. 3. A similar study has also been conducted in [34] which the results of [33] have been verified by discrete elements instead of CMOS implementation.

In the proposed structure in [33], using a sampling circuit with a sampling period of $T$ seconds, the input is sampled periodically. Using cascaded samplers, required delays are implemented. Instead of resistors which are not configurable, memristors (as configurable resistors) are used to implement coefficients. Thus, the filter can be adapted to any form by setting memristors' conductance to different values, thus increasing the configurability of filters. The required addition operation can be implemented by operational amplifiers (Op-Amps) in their small signal region.

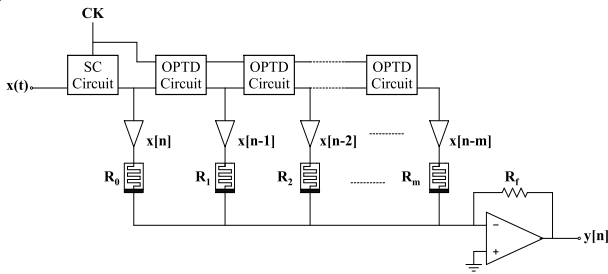

**Fig. 3** *The whole FIR filter proposed in [33]*

The proposed structure in [33] which is the base of our research does not have the capability to implement negative coefficients which they are necessary in implementing Type III and Type IV FIR filters. Moreover, its sampling circuit which is based on Switched–Capacitor topology has limitations in accepting a wide range of input voltages. Furthermore, the OPTD circuit which is responsible for handling delays is not suitable for high orders of FIR filters, because of its complexity and inappropriate cascading technique.

### 1.3. *Current Work*

The purpose of this paper is proposing a novel structure to implement configurable FIR filters using analog circuit elements and memristors and demonstrate its accuracy and proper operation. As the main contribution of this work, a new approach for implementing negative coefficients has been proposed; a heuristic method to reduce the error of implemented coefficients has been applied. Furthermore, using appropriate circuits as the sampling blocks, the input range has been increased. The simple delay units and their interconnection scheme have enabled our proposed structure to avail higher orders of filters compared to other works.

The rest of this paper is organized as the following. The proposed circuits have been provided in section 2. Simulations and their results are all covered in section 3. Section 4 is dedicated to the conclusion.

## 2. Main research

The proposed structure for an m$th$-order memristor-based FIR filter is shown in Fig. 4. In this scheme, the input signal is sampled by the track-and-hold unit. Passing through delay units, the delayed samples ($x[n-m]$) are produced. The scaling buffers ($a$) are used to attenuate the input samples if their amplitude is more than the dead-zone of the memristors (0.1V specified in [34, 35]). Multiplication of the samples by the coefficients and adding are done by the memristors and op-amps, simultaneously. Each block of the circuit is explained in the following and the advantages of our work over other implementations have been discussed.

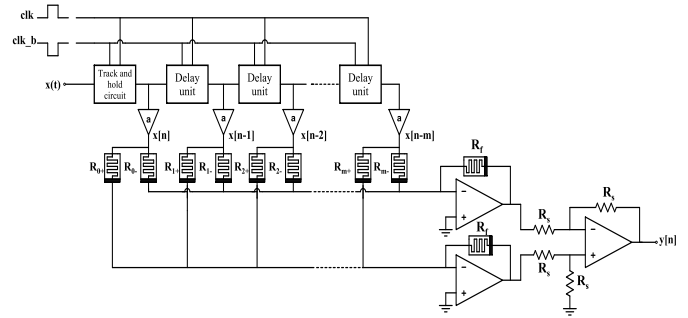

**Fig. 4** *The proposed memristor-based FIR filter(dashed area is for handling negative coefficients)*

### 2.1 *Handling Negative Coefficients*

As mentioned before, the proposed memristor-based FIR filter in [33] is only suitable for implementing specific filters which their coefficients are all positive; because memristors could not get negative values. As the main contribution of this article, a new circuit topology (Fig. 5) to handle negative coefficients is proposed.

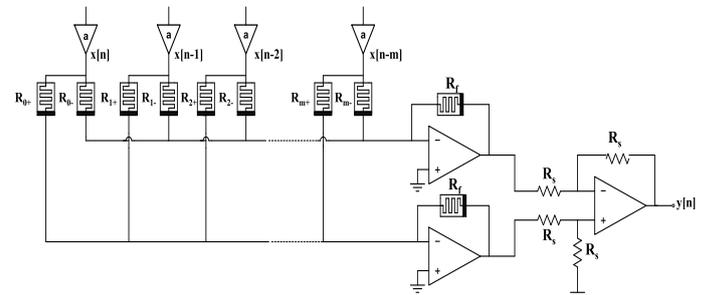

**Fig. 5** *Handling negative coefficients*

Applying circuit analysis fundamentals and using (1), output of the proposed structure is determined by (2):

$$y[n] = \left(\frac{R_f}{R_{0+}} - \frac{R_f}{R_{0-}}\right)x[n] + \left(\frac{R_f}{R_{1+}} - \frac{R_f}{R_{1-}}\right)x[n-1] +$$
$$\cdots + \left(\frac{R_f}{R_{m+}} - \frac{R_f}{R_{m-}}\right)x[n-m] \qquad (2)$$

where the filter coefficients $b_i$ are described by (3):

$$b_i = \frac{R_f}{R_{i+}} - \frac{R_f}{R_{i-}} ; \quad i = 0, 1, \ldots, m \tag{3}$$

In which $R_{i+}$ and $R_{i-}$ are memristance of memristors for handling positive and negative numbers, respectively. In other words, to make a coefficient negative, it is sufficient that $R_{i+} > R_{i-}$, and vice versa. $R_f$ can be used as a degree of freedom to assign the range of memristance.

The maximum resolution of setting memristors is about 7 to 8 bits [34, 35]. In this case, the range of memristance would be divided into about 128 or 256 values. Therefore, in setting filter coefficients, memristance values will be set to one of the acceptable values limited by resolution which this will lead to unavoidable discretization errors in coefficients' values. The following methods are recommended for calculating the value of memristors in the filter.

### 2.1.1. Simple method:
According to the determined range for memristance [34, 35], for building positive coefficients, $R_{i-}$ should be set to $R_{m-max}$ which is the maximum boundary of memristance. Therefore the negative term of (3) would be minimized. $R_f$ should be chosen in a way that all $R_{i+}$ fall within the determined range for memristance. For building negative coefficients, $R_{i+}$ should be set to $R_{m-max}$ and the same approach should be applied.

It is obvious that having only one degree of freedom ($R_f$), it is difficult to set the calculated $R_{i+}$ or $R_{i-}$ to the memristance which is limited by resolution of memristors or tuning circuits.

### 2.1.2. Advanced method:
In order to decrease the error of the calculated filter coefficients, $R_f$, $R_{i+}$, $R_{i-}$ values are determined applying a heuristic approach; hence, there are three degrees of freedom which can result to a better accuracy in setting coefficients $b_i$. The maximum likelihood criteria is used to minimize the following function in a way that the memristance forming each $\overline{b_i}$ are selected considering the acceptable values limited by resolution and giving the minimum F:

$$F = \sum_{i=0}^{m} (\, b_i - \overline{b_i}) \tag{4}$$

where $b_i$ is the target value (the ideal value of the coefficient) and $\overline{b_i}$ is the erroneous calculated value of the corresponding coefficient (the value which is implementable with memristors and suggested by the heuristic search algorithm).

It is worth to mention that the op-amp on the final stage is responsible for subtracting the results of the first stage op-amps; the value of all the final stage resistors have been selected to be equal deliberately here to give the unit gain to its inputs. Changing these resistors ($R_s$) can give more degrees of freedom in selecting memristors' values, and also amplification or attenuation of the filtered signal.

## 2.2. Sampling Circuit

Sampler as the first stage of the filtering operation can be implemented using multiple topologies. The most well-known sampling circuit utilizes an N-MOS switch which is illustrated in Fig. 6(a). The on-resistance of the switch versus input voltage is shown in Fig. 6(b). It is clear that as the input voltage becomes bigger, on-resistance of the switch becomes worse. This topology has been used in [33].

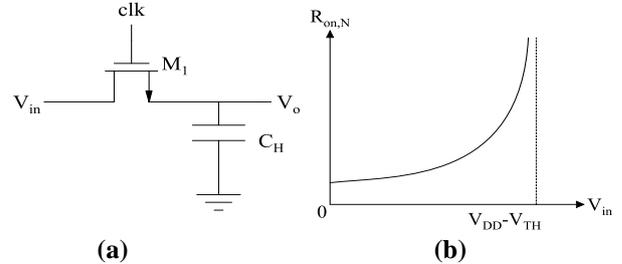

**Fig. 6** *N-MOS Sampling Circuit*

(a) N-MOS Switch (b) $R_{on}(on)$ versus $V_{in}$

Another approach is using a P-MOS switch instead of a N-MOS one in the sampling circuit. This is illustrated in Fig. 7(a) and its on resistance versus input voltage characteristic is shown in Fig. 7(b). Unlike a N-MOS switch, a P-MOS switch shows good switching features in high voltages while it cannot sample the input correctly for low voltages.

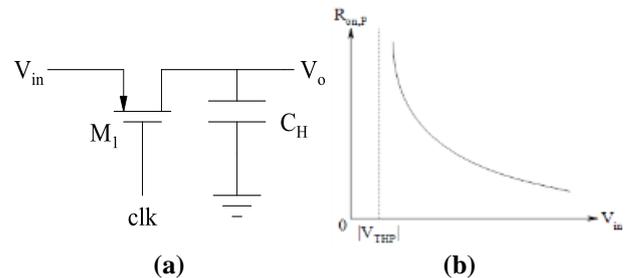

**Fig. 7** *P-MOS sampling circuit*

(a) P-MOS Switch (b) $R_{DS}(on)$ versus $V_{in}$

According to the deficiency of N-MOS and P-MOS switches, for having a larger range of input voltage, a boot-strap switch driving circuit can be used [36] to stabilize the Gate-Source voltage of the switch. Using boot-strap driver increases the overhead. Therefore, a complementary switch which includes both N-MOS and P-MOS switches in parallel without any boot-strap gate drivers can be used. This structure, called a CMOS switch, is shown in Fig. 8(a) and its characteristic is shown in Fig. 8 (b). Obviously, for a significantly wide range of input voltage, the complementary switch shows a smooth and approximately linear on-resistance. It has to be mentioned that for driving this switch, it is necessary to have a non-overlapping clock generator to produce clk and clk_b illustrated in Fig. 4 [37, 38] . While overcoming the on-resistance problem of the sampling circuit, other side effects such as clock feed-through [39, 40], kT/C noise [41, 42], and charge injection [43, 44]should not be neglected.

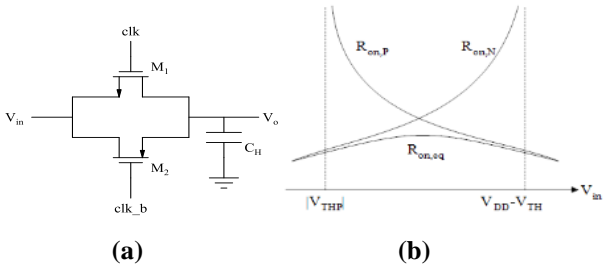

**(a)**                **(b)**

***Fig. 8** CMOS sampling circuit*

**(a)** CMOS Switch **(b)** $R_{DS}(on)$ versus $V_{in}$

In our proposed structure for the sampler, in order to overcome the charge injection and the clock feed-through problem, Dummy switches have been used. The overall sampling circuit is presented in Fig. 9.

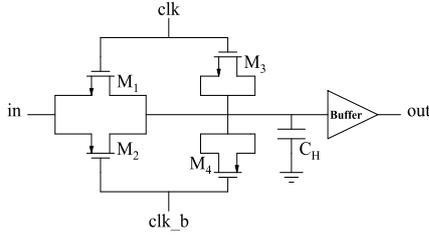

***Fig. 9** Sampling circuit*

### 2.3. Delay Circuit

The designed circuit in [33] uses cascaded mono-stable circuits for implementing delay units which it not only requires extra hardware but it also limits the flexibility of the circuits to be used in high order filters. In order to make our design suitable for any order, a Master-Slave topology has been proposed which is shown in Fig. 10. In this circuit, the first stage which receives the input voltage is called the Master, and similarly the second stage is called the Slave. When *clk* becomes *low*, master switches will be turned on and the output of the master ($V_x$) will be almost equal to the input voltage. During *low* state of *clk,* the slave switches remain off. After *clk* goes to the *high* state, the slave switches will be turned on and $V_x$ will be mapped to the output. Similarly, during the *high* state of *clk*, master switches remain inactive in order not to change the output in the same clock level. Using this structure, the control circuit presented in [33] is omitted; thus, hardware area is reduced. Clock feed-through is also handled. It is significant to note that *clk* and *clk_b* must be completely non-overlapping; in other words, *clk_b* is the *logical NOT* of *clk*.

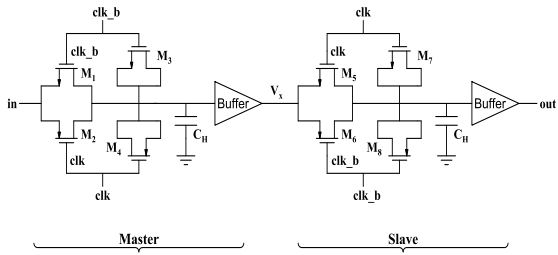

***Fig. 10** Master-Slave delay unit.*

### 2.4. Utilized Memristor Model

The model for the memristors used in this paper is the HP model [9]. Ordinary switching characteristics of this model are shown in Fig. 11, which has been determined by sweeping the voltage across the memristor in [35]. To better illustrate the range of resistors, approximate numbers for the memristance of the device being tested have been presented in Fig. 11. According to the stability test of the memristor, after it was adjusted to a certain value, it retained its value for 50,000 seconds while being applied a voltage about 100 mV. This test was conducted in 350 K degrees and the value of the memristor was read once every two seconds. Hence, in this work, if the memristors are used in ambient temperature and it is guaranteed that the value of the voltage applied to the memristor not exceed 100mV, it would be possible for memristors to retain their set values more than ten years [35].

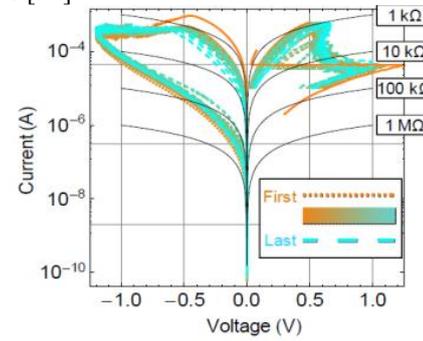

***Fig. 11** Switching characteristics of the memristor* [35]

The calculated value for $R_m$ and $R_f$ should be adjusted to the memristor by tuning circuits [33, 44]. Having tuning circuits for each memristor can lead to the programmability of the filter for different applications. As mentioned before, it must be noted that a scaling stage (*a*) is used as a scalar to keep the voltage applied across the memristor below 100mV. This will guarantee that memristors' values for a specific FIR filter will not change by the input signal [34].

## 3. Simulation Results and Discussion

In this section, the simulation results of the memristor-based filter have been presented. All the simulations have been done within the Cadence environment. First, the correct operation of the sampler unit and delay units has been investigated. Then, the structure for a low-pass and a high-pass filter is tested and the results have been illustrated.

To compare the accuracy of the two methods mentioned in section 2.1, they have been used in determining coefficients for 6, 7, and 8 bits of resolution in the simulations. For simplicity, only the output signal for 7-bit resolution has been presented.

### 3.1. Samplers and Delay Unit

The sinusoidal signal defined by (5), represented in blue in Fig. 12, is applied to the circuit shown in Fig. 4. The circuit being tested in this section is a fifth-order filter. $C_H$ is set to 5 pF. The output of the sampler circuit is illustrated by black. The other colors represent delayed samples. Fig. 12 verifies the proper operation of the delay units.

$$x(t) = 0.2 \sin(2\pi * 2 * 10^3 \, t) \qquad (5)$$

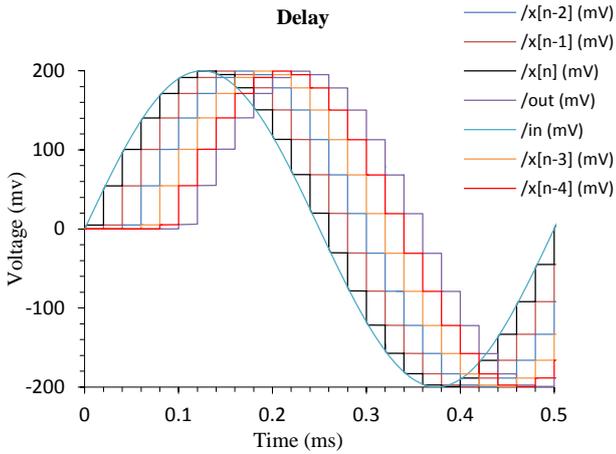

**Fig. 12** *The delayed versions of the input for five orders*

### 3.2. Implementation of a Low-Pass Filter

A low-pass FIR filter with the properties presented in Table 1 is aimed to be designed. The determined coefficients are calculated using MATLAB FDA Toolbox. It is to be mentioned that in the majority of laboratory implemented memristors there is an acceptable range for their memristance. For instance, in Fig. 11, it can be seen that the minimum memristance is $1 K\Omega$ but its maximum boundary is $1 M\Omega$. Therefore, $R_f$ should be calculated and selected in a way to make sure that $R_{i+}$ and $R_{i-}$ fall in their acceptable region.

**Table 1** Characteristics of the low-pass filter

| Filter Type | $f_s$ | $f_c$ | Order |
|---|---|---|---|
| Low-pass | $400 \, KHz$ | $20 KHz$ | 16 |

To test the designed circuit, a signal represented by equation (6) is applied to the filter which contains two different frequency components with equal amplitudes. According to Table 1, the sampling frequency ($f_s$) is 400kHz. The cut-frequency ($f_c$) is set to 20 kHz.

$$x(t) = 0.4 \sin(2\pi * 5 * 10^3 \, t) + 0.4 \sin(2\pi * 60 * 10^3 \, t) \qquad (6)$$

For the designed filter, coefficients have been calculated using both simple and advanced methods based on 6, 7, and 8 bits of resolution. To avoid mass of data, target and calculated values for memristors and coefficients have been presented in Table 2 for only 7-bit resolution.

The percentage of error between the target coefficient value and its calculated value is obtained using equation (7).

$$Error_{bi} = abs(\frac{bi_{target} - bi_{calculated}}{bi_{target}}) \times 100 \qquad (7)$$

The percentage of error for the coefficients has been calculated and visualized in Fig. 13 for 6, 7, and 8 bits of resolution. It is clear that the maximum error occurred in setting coefficients for advanced method is less that 1% while it is more than 40% for the simple method.

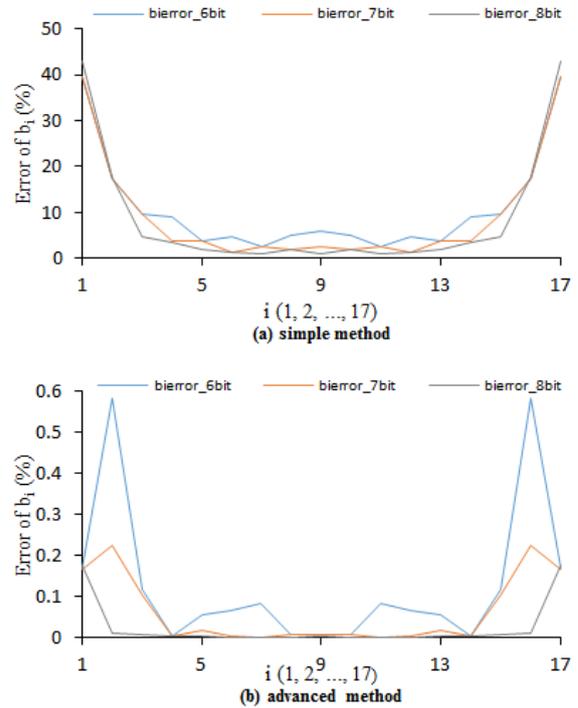

**Fig. 13** *Coefficients error for low-pass filters in 6, 7, and 8 bits of resolution*

**(a)** Simple method **(b)** Advanced method

Fig. 14 illustrates the frequency response of the low-pass filter based on simple and advanced method compared to the response of the ideal (original) filter. It demonstrates the high accuracy and usefulness of the advanced method in contrast to the simple method for calculation and optimization of the value of memristors thus coefficients.

As an example, Fig. 15 shows the output of the circuit for both simple and advanced methods; the yellow trace (input signal) contains 5 kHz and 60 kHz components while, as expected, the output is just the shifted version of the low-frequency component of $x(t)$. The shift between input and output is the intrinsic characteristic of FIR filters. The outputs for both simple and advanced methods are similar which this is the result of the fact that the significant difference between two methods is mainly in two coefficients which they cannot change the characteristic of the filter considerably.

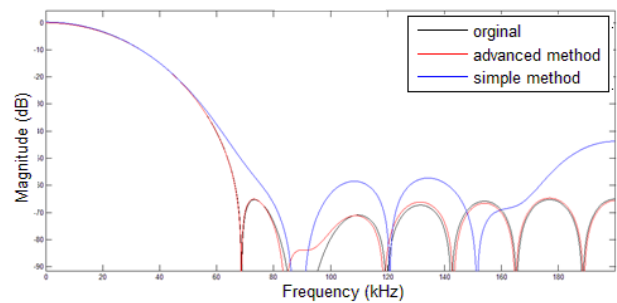

**Fig. 14** *Frequency response of the designed low-pass filter*

| Coefficient | Target $b_i$ | Simple method ($R_f$ = 136 K$\Omega$) | | | Advanced method ($R_f$ = 48 K$\Omega$) | | |
|---|---|---|---|---|---|---|---|
| | | Calculated $b_i$ | $R_{i+}(K\Omega)$ | $R_{i-}(K\Omega)$ | Calculated $b_i$ | $R_{i+}(K\Omega)$ | $R_{i-}(K\Omega)$ |
| b0 | 0.00154566 | 0.002156537 | 984 | 1000 | 0.001543060 | 688 | 703 |
| b1 | 0.00569446 | 0.006681507 | 953 | 1000 | 0.005681641 | 735 | 805 |
| b2 | 0.01524095 | 0.016683004 | 891 | 1000 | 0.015224887 | 516 | 618 |
| b3 | 0.03180934 | 0.032968697 | 805 | 1000 | 0.031811106 | 501 | 750 |
| b4 | 0.05533191 | 0.05734059 | 703 | 1000 | 0.055341896 | 415 | 797 |
| b5 | 0.08306805 | 0.08421784 | 618 | 1000 | 0.083064799 | 259 | 469 |
| b6 | 0.10980299 | 0.112479831 | 547 | 1000 | 0.109803747 | 204 | 383 |
| b7 | 0.12928842 | 0.131556061 | 508 | 1000 | 0.129297782 | 110 | 157 |
| b8 | 0.13643644 | 0.140032665 | 493 | 1000 | 0.136423789 | 188 | 407 |
| b9 | 0.12928842 | 0.131556061 | 508 | 1000 | 0.129297782 | 110 | 157 |
| b10 | 0.10980299 | 0.112479831 | 547 | 1000 | 0.109803747 | 204 | 383 |
| b11 | 0.08306805 | 0.08421784 | 618 | 1000 | 0.083064799 | 259 | 469 |
| b12 | 0.05533191 | 0.05734059 | 703 | 1000 | 0.055341896 | 415 | 797 |
| b13 | 0.03180934 | 0.032968697 | 805 | 1000 | 0.031811106 | 501 | 750 |
| b14 | 0.01524095 | 0.016683004 | 891 | 1000 | 0.015224887 | 516 | 618 |
| b15 | 0.00569446 | 0.006681507 | 953 | 1000 | 0.005681641 | 735 | 805 |
| b16 | 0.00154566 | 0.002156537 | 984 | 1000 | 0.001543060 | 688 | 703 |

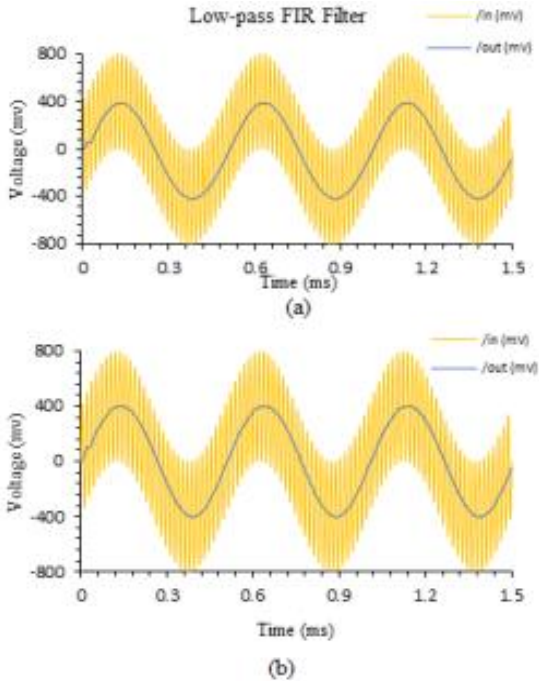

**Fig. 15** *Input/Output of the low-pass filter*

**(a)** simple method **(b)** advanced method

### 3.3 Implementation of a High-Pass Filter

Similar to the design process of the low-pass filter, a high-pass one is designed and evaluated to show the feasibility of negative coefficients which has been proposed in this work. The requirements of the 11th order filter are shown in Table 3. The calculated values for the memristors are presented in Table 4.

**Table 3.** Characteristics of the high-pass filter.

| Filter Type | $f_s$ | $f_c$ | Order |
|---|---|---|---|
| Highpass | 500 KHz | 10KHz | 11 |

The applied signal to the filter is described by equation (8) which consists of two frequency components, 2 kHz and 90 kHz. To illustrate the operation of the filter, the amplitude of the high-frequency component of the input signal is intended to be much smaller than the low-frequency component. The simulation result in Fig. 16 shows that the high- frequency component of the input signal exists in the output and the low-frequency component is totally filtered.

$$x(t) = 0.12 \sin(2\pi * 2 * 10^3\ t) + 0.03\ sin\ (2\pi * 90 * 10^3\ t) \qquad (8)$$

It is also visible in Fig. 17 that the maximum error for the simple method, like the low-pass filter, is extremely high (about 70%). In contrast, the error for advanced method is below 1% which shows the advanced method's competency over the simple method. This can also be justified using Fig. 16. In Fig. 16(a) the low pass component's smooth existence is not ignorable in the output while in Fig. 16(b) the output does not seem to have any considerable low-frequency component. Similar to the low-pass filter in section 3.2, the frequency response of the designed high-pass filter is shown in Fig. 18 which demonstrates that the performance of the advanced method for calculation and optimization of filter coefficients is slightly better than the simple method.

**Table 4** Values of target and calculated coefficients and memristors of the high-pass filter for 7 bit resolution

| Coefficient | Target $b_i$ | Simple method ($R_f$ = 624 KΩ) | | | Advanced method ($R_f$ = 243 KΩ) | | |
|---|---|---|---|---|---|---|---|
| | | Calculated $b_i$ | $R_{i+}(K\Omega)$ | $R_{i-}(K\Omega)$ | Calculated $b_i$ | $R_{i+}(K\Omega)$ | $R_{i-}(K\Omega)$ |
| b0 | 0.00341238 | -0.002506024 | 1000 | 996 | 0.003404031 | 742 | 750 |
| b1 | 0.01072902 | -0.014689867 | 1000 | 977 | 0.010744041 | 586 | 602 |
| b2 | 0.01438623 | -0.03631746 | 1000 | 945 | 0.014369824 | 618 | 641 |
| b3 | -0.01139188 | -0.0763367 | 1000 | 891 | -0.011372795 | 719 | 696 |
| b4 | -0.11947815 | -0.18220155 | 1000 | 774 | -0.119496585 | 415 | 344 |
| b5 | -0.6108322 | -0.641720081 | 1000 | 493 | -0.610829511 | 969 | 282 |
| b6 | 0.6108322 | 0.641720081 | 493 | 1000 | 0.610829511 | 282 | 969 |
| b7 | 0.11947815 | 0.18220155 | 774 | 1000 | 0.119496585 | 344 | 415 |
| b8 | 0.01139188 | 0.0763367 | 891 | 1000 | 0.011372795 | 696 | 719 |
| b9 | -0.01438623 | 0.03631746 | 945 | 1000 | -0.014369824 | 641 | 618 |
| b10 | -0.01072902 | 0.014689867 | 977 | 1000 | -0.010744041 | 602 | 586 |
| b11 | -0.00341238 | 0.002506024 | 996 | 1000 | -0.003404031 | 750 | 742 |

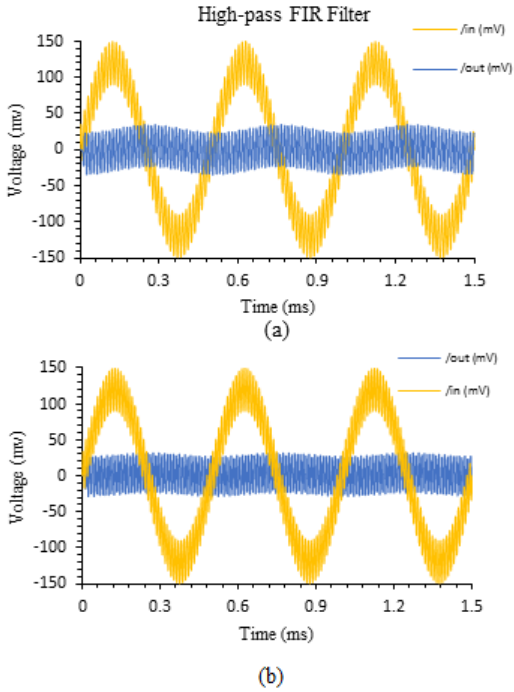

**Fig. 16** *Input/Output of the high-pass filter*

(**a**) simple method (**b**) advanced method

## 4. Conclusion

Being a fundamental part of signal processing applications, FIR filters have too much cost in terms of energy for processors while they can be implemented using analog components. In this work, a new scheme is proposed to implement general purpose configurable memristor-based FIR filters. A new circuit for handling negative filter coefficients is developed and tested. In addition, proposing a new technique for sampling and delay units, the problems of the former research, namely, bad sampling switch selection, clock feed-through, limited order of the filter have been overcome. Finally, to verify the propoer operation of the designed circuits, a low-pass and high-pass filter is simulated. To enhance the accuracy of implemented coefficients, a heuristic method is applied to calculating memristors' values. The simulations show correct operation of the proposed structures and improvement in coefficients accuracy.

The potential future research can be implementing other configurable filters along with applying other heuristic or nature-inspired optimization algorithms to calculating memristor values especially in high-order filters to save both time and enhance accuracy.

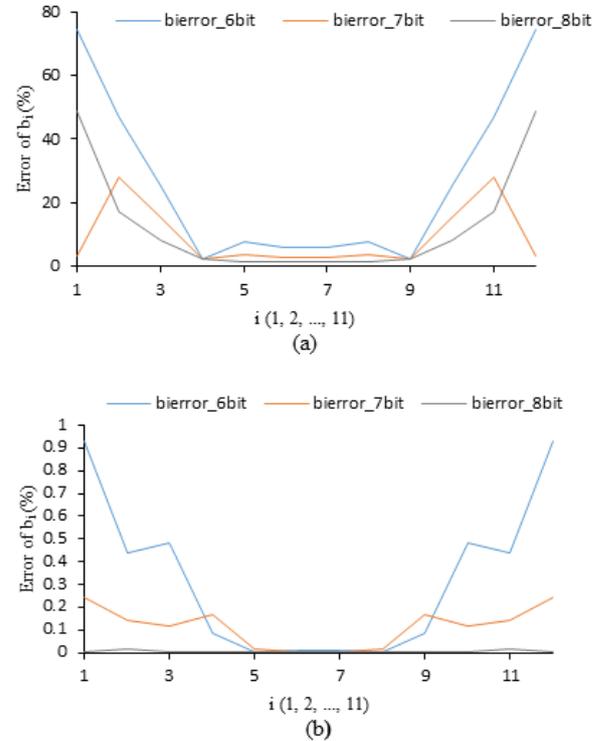

**Fig. 17** *Coefficients' error for the high-pass filter in 6, 7, and 8 bits of resolution*

(**a**) Simple method (**b**) Advanced method

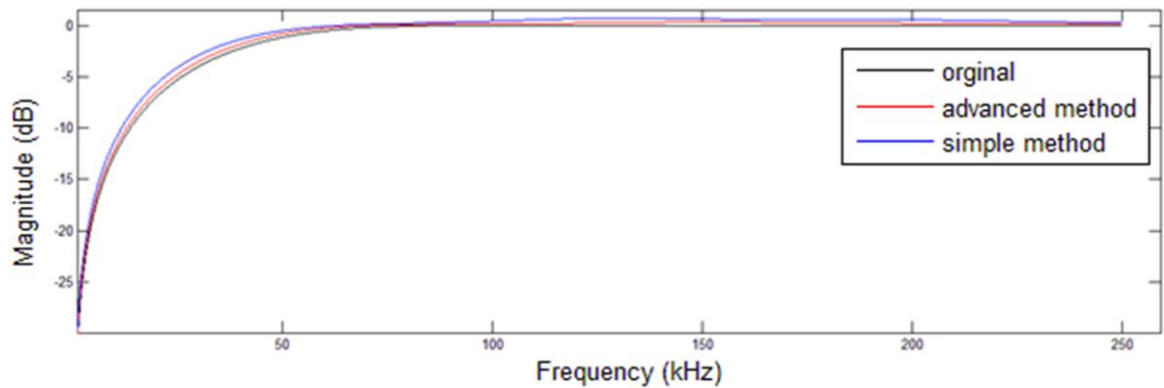

**Fig. 18** *Frequency response of the designed high-pass filter*

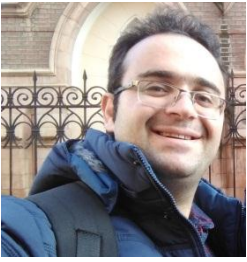

**Mohammad Hemmati** received his B.Sc and M.Sc degree from University Of Zanjan, Zanjan, Iran in Electrical and Electronics Engineering, in 2013 and 2016, respectively.

He was a visiting lecturer at Electrical Engineering department at University of Zanjan, teaching Digital Design and Microprocessors Laboratories for more than two years. He is currently working as an embedded systems designer and CEO at Ofogh Sanaat Paydar Zangan doing research on biomedical signal processing, embedded systems, and computer vision. His research interests include Embedded Systems, FPGA-based Digital Design, Geometry of Vision, Fault-Tolerant Systems, and Artificial Intelligence in Biomedical applications.(Email: m_hemmati@znu.ac.ir )

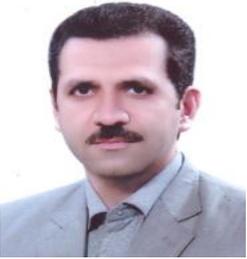

**Vahid Rashtchi** was born in 1967 in Zanjan, Iran. He received his B.Sc. in electrical engineering from Tabriz University in 1991 and his M.Sc. and Ph.D. in electrical engineering from Amirkabir University of Technology, Tehran, Iran, in 1993 and 2001, respectively.

He is currently working as an associate professor at University of Zanjan, working on power electronics and applications of artificial intelligent systems since 2001. Up to now, he has had several positions such as "Head of the Engineering Faculty", "Research Vice Chancellor" at University of Zanjan and "Chairman of the National Elites Foundation- Zanjan Branch". (Email: rashtchi@znu.ac.ir )

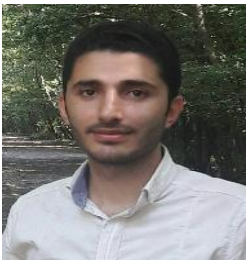

**Ahmad Maleki (Non-member)** was born in 1992 in Lorestan, Iran. He received his B.Sc. in electrical engineering from Urmia University, Urmia, Iran, in 2014. He holds a Masters in Electronics from University of Zanjan, Zanjan, Iran doing research on Memristors and their applications.

His research interests include FPGA-based Design, signal processing, and image processing. (Email: maleki.a@znu.ac.ir)

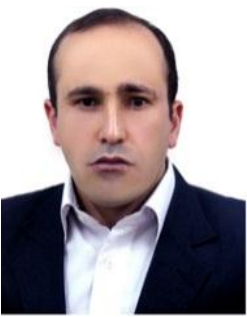

Siroos Toofan **(Non-member)** received the B.Sc. degree in Electronics Engineering from Amirkabir University of Technology (Tehran Polytechnic) in 1999, and the MSc. And PhD degree in Electronics Engineering form Iran University of Science and Technology (IUST) in 2002 and 2008, respectively. During 2007 to 2008, on his sabbatical leave, he was with the VLSI group of Politechnico di Torino and in the Microelectronics- Integrated Circuits Lab. of the Politechnico di Milano in Italy. He has been working as an assistant professor with the Department of Electrical Engineering, University of Zanjan, since 2009. His current research activities include the design of CMOS Analog Integrated Circuits, RF Integrated Circuits and Capacitive Sensors Readout Circuits. (Email: s.toofan@znu.ac.ir)